\documentclass[final,3p,times,sort&compress]{elsarticle}

\usepackage{amssymb}

\usepackage{booktabs}
\usepackage{multirow}
\usepackage{array}
\usepackage{makecell}

\usepackage{hyperref}

\usepackage{soul,color,xcolor}

\usepackage{pdflscape}
\usepackage{afterpage}

\newcommand{\We}{\color{black}{W\!e}}%
\newcommand{\Oh}{\color{black}{Oh}}%
\usepackage{amsmath}
\usepackage{lineno}

\journal{Fuel}

\usepackage{nomencl} 

\renewcommand\nomgroup[1]{%
  \item[\bfseries
  \ifstrequal{#1}{G}{Greek symbols}{%
  \ifstrequal{#1}{S}{Subscripts}{}}%
]}

\makenomenclature

\begin{document}

\begin{frontmatter}

\title{Transitions of breakup regimes for viscous droplets in airflow}
\author{Zhikun Xu}
\author{Tianyou Wang}
\author{Zhizhao Che\corref{cor1}}
\ead{chezhizhao@tju.edu.cn}
\cortext[cor1]{Corresponding author}
\address{State Key Laboratory of Engines, Tianjin University, Tianjin, 300350, China.}

\begin{abstract}
Droplet breakup is important in many natural and industrial processes, but the current classification of breakup regimes, especially the intermediate breakup regime, is ambiguous. In this study, the transitions of breakup regimes for viscous droplets are investigated experimentally using high-speed imaging taken from a side view and a 45$^{\circ}$ view. Based on the morphology change in the middle of the droplet, the breakup regimes are classified into no-breakup, bag breakup, bag-stamen, low-order multimode, high-order multimode, and shear-stripping breakup. The droplet morphologies in different regimes and the corresponding transitions are discussed in detail. The droplet viscosity dissipates the kinetic energy transferred by the airflow during the initial droplet flattening, and affects the development of the Rayleigh-Taylor instability wave after the flattening. Through the analysis of the droplet deformation and the Rayleigh-Taylor instability with the droplet viscosity taken into account, the transition conditions of different regimes are obtained in a regime map. By further considering the relative velocity loss between the droplet and the airflow, the ranges of the dual-bag breakup in the low-order multimode regime and the droplet retraction in the bag-stamen regime are determined.
\end{abstract}


\end{frontmatter}

\section{Introduction}\label{sec:sec1}

With the requirement of carbon neutralization and the development of energy conversion systems, the sources and types of fuels are greatly expanded, such as biofuels \citep{Ravikrishna2021Surrogatefuels,Wandel2021EmulsifiedBiofuel,Nguyen2020biodiesels}, gelled propellants \citep{Manisha2021Gel, Yoon2017Gel}, and emulsified oils \citep{Gavaisesb2020Emulsions, Kim2019Emulsifiedoils}. A common feature of these fuels is their higher viscosity compared to conventional diesel or gasoline, which hence requires higher energy for fuel atomization. In addition, the atomization of viscous liquids is also a fundamental issue in various applications of machinery, chemical engineering, pharmaceutical manufacturing, and food engineering, such as particle manufacturing, spray drying, and spray coating. Liquid atomization generally involves the primary breakup of a liquid jet and the subsequent secondary breakup of droplets. The secondary breakup controls the size distribution of the final droplets and affects the heat/mass transfer, which plays a fundamental role in liquid atomization \citep{Lefebvre2017Atomization, Michele2019SpraySystems,Dai2001MultimodeBreakup}. Hence, the study of viscous droplet breakup is important for understanding and improving the atomization processes.

\begin{table}[]
\label{tab:my-table}
\begin{tabular}{|p{1cm}p{6.28cm}p{1cm}p{6.28cm}|}
\hline
\multicolumn{2}{|l}{\textbf{Nomenclature}}           & $\rho$  & density                      \\
                        &                           & $\sigma$  & surface tension              \\
$a$           & acceleration                        & $\lambda$  & wavelength                   \\
$C$           & constant                            & \multicolumn{2}{l|}{\textit{\textbf{Subscripts}}}  \\
$d$           & droplet diameter                    & $c$        & critical value               \\
$F$           & force                               & $d$        & droplet                      \\
$l$           & length                              & $D$        & drag                         \\
$\Oh$         & Ohnesorge number, $\Oh={\mu}/\sqrt{{\rho}{d}\sigma}$   & $g$        & gas                          \\
$t$           & time                                & $i$        & initial value for flattening \\
$t^*$         & characteristic time                 & local      & local value                  \\
$T$           & dimensionless time, $T=t/{t^*}$                  & $RT$       & Rayleigh--Taylor instability  \\
$u$           & velocity                            & $t$        & droplet thickness            \\
$\We$         & Weber number, ${\We}={\rho}u^2{d}/\sigma $                        & $w$        & droplet width                \\
$\mu$         & viscosity                           & 0          & initial or inviscid value \\
\hline
\end{tabular}%
\end{table}

There have been many studies on the breakup of droplets under the action of airflow. According to the droplet morphology, the droplet breakup can be classified into three main breakup regimes, i.e., bag breakup, shear-stripping breakup, and intermediate breakup regimes \citep{Gavaises2016ModerateWe,Guildenbecher2009SecondaryAtomization,Hsiang1995BreakupCategory}. In the bag breakup regime, the droplet first deforms and is stretched into a thin hollow bag attached to a thicker toroidal rim, which is caused by the piercing from the droplet center after the droplet is flattened \citep{Kulkarni2014BagBreakup, Zhao2011BagBreakup}. The mechanism governing the bag breakup is the development of Rayleigh-Taylor (RT) instability waves at the interface \citep{Theofanous2011DropBreakup, Zhao2010BagBreakup} or the formation of Hill vortices inside the droplet\citep{Sojka2021MultimodeBreakup, Rimbert2020DropletDeformation}. In the shear-stripping breakup regime, the main feature of the droplet morphology is the appearance of Kelvin-Helmholtz (KH) instability waves at the periphery of the droplet, which lead to continuous mass stripping \citep{Wang2020MachNumber, Sharma2021Aerobreakup, Theofanous2012ViscousLiquids}. The intermediate breakup regime generally consists of several sub-regimes and has the morphological features of both the bag breakup and the shear-stripping breakup \citep{Gavaises2016ModerateWe,Radhakrishna2021HighOh}. The different breakup regimes are determined mainly by the Weber number (${\We_g} = {\rho _g}u_g^2{d_0}/\sigma $) and the Ohnesorge number ($\Oh = {\mu _d}/\sqrt {{\rho _d}{d_0}\sigma}$), where ${\rho _g}$ is the gas density, ${\rho _d}$ is the droplet density, ${u_g}$ is the gas velocity, ${d_0}$ is the initial droplet diameter, and $\sigma $ is the droplet surface tension. The Weber number $\We_g$ and the Onesorge number $\Oh$ are measures of the ratios of the aerodynamic force and the viscous force to the surface tension force, respectively \citep{Guildenbecher2009SecondaryAtomization}. The transitional $\We_g$ between different regimes increases with the increase in $\Oh$ \citep{Hsiang1995BreakupCategory, Theofanous2012ViscousLiquids}.

Since the fragment size distribution and the breakup time are remarkably different for different breakup regimes \citep{Pilch1987BreakupTime, Jackiw2022SizeDistribution, Guildenbecher2017FragmentSize}, it is important to accurately classify the breakup regimes. Different studies have formed a basic consensus on the transition of different breakup regimes, but these studies mainly focus on the bag breakup \citep{Zhao2011BagBreakup, Theofanous2012ViscousLiquids, Yang2017TransitionsWeberNumber} and the shear-stripping breakup \citep{Theofanous2012ViscousLiquids, Zhao2018Instibility}. For the intermediate breakup regime, there is still debate on the detailed morphology and the exact transitional ${\We_g}$. Table \ref{tab:tab1} shows the classification of breakup regimes and the corresponding $\We_g$ ranges when $\Oh < 0.1$ in different studies. Different researchers classified the intermediate breakup regime into different sub-regimes, and the ranges of different sub-regimes overlap and are inconsistent. This discrepancy is not only related to the differences in the experimental techniques and the experimental uncertainty \citep{Gavaises2016ModerateWe}, but also to the incomplete morphology obtained due to limited experimental conditions (often lacking information on the windward side of the droplet) \citep{Theofanous2011DropBreakup, Sharma2021Aerobreakup}. Moreover, for low $\Oh$, the morphology of the intermediate regime is complex and changes rapidly with the change of $\We_g$, which may be the main reason for the discrepancy. As for high $\Oh$, it can be inferred that due to viscous dissipation, the change in the morphology of the intermediate regime slows down and the sub-regimes can be more easily distinguished. However, the study on the transitions of the intermediate regime of the viscous droplet is inadequate. Krzeczkowski \citep{Krzeczkowski1980ViscosityBreakup} and Faeth \emph{et al.}\ \citep{Hsiang1992SecondaryBreakup, Hsiang1995BreakupCategory} analyzed the transitions of the different regimes in a $\We_g$--$\Oh$ plot. They simply distinguished the intermediate regime as the multimode breakup. However, in subsequent studies, more complex morphologies were identified in the intermediate regime, such as bag-stamen \citep{Zhao2013BagStamen}, dual-bag \citep{Cao2007DualbagBreakup}, rebound \citep{Theofanous2012ViscousLiquids}, and wings \citep{Radhakrishna2021HighOh}. These discrepancies indicate that further study is needed on the transition of the breakup regimes for viscous droplets.

\begin{table}[]
\caption{Droplet breakup regimes under different $\We_g$ at low $\Oh$.}
\label{tab:tab1}
\resizebox{\columnwidth}{!}{%
\begin{tabular}{|c|c|c|c|c|c|}
\Xhline{4\arrayrulewidth}
& Krzeczkowski \citep{Krzeczkowski1980ViscosityBreakup} & Dai and Faeth \citep{Dai2001MultimodeBreakup} & Zhao \emph{et al.}\ \citep{Zhao2010BagBreakup} & Sojka \emph{et al.}\ \citep{Guildenbecher2017FragmentSize} & Jain \emph{et al.}\ \citep{Jain2015SecondaryBreakup}\\
\Xhline{4\arrayrulewidth}
Bag breakup                           & $10<\We_g<18$                                                                      & $13<\We_g<18$                                                                          & $12<\We_g<16$                                                                        & $9<\We_g<15$                                                                           & $12<\We_g<16$                                                        \\ \hline
\multirow[c]{7}{*}{Intermediate breakup} & \multirow[c]{3}{*}{\begin{tabular}[c]{@{}c@{}}$18<\We_g<30$\\ (Bag-jet)\end{tabular}}   & \multirow[c]{3}{*}{\begin{tabular}[c]{@{}c@{}}$18<\We_g<40$\\ (Bag-plume)\end{tabular}}   & \begin{tabular}[c]{@{}c@{}}$16<\We_g<28$\\ (Bag-stamen)\end{tabular}                 & \multirow[c]{3}{*}{\begin{tabular}[c]{@{}c@{}}$15<\We_g<31$\\ (Multimode)\end{tabular}}   & \begin{tabular}[c]{@{}c@{}}$16<\We_g<24$\\ (Bag-stamen)\end{tabular} \\ \cline{4-4} \cline{6-6}
                                      &                                                                                    &                                                                                      & \begin{tabular}[c]{@{}c@{}}$28<\We_g<41$\\ (Dual-bag)\end{tabular}                   &                                                                                      & \begin{tabular}[c]{@{}c@{}}$24<\We_g<45$\\ (Dual-bag)\end{tabular}   \\ \cline{2-6}
                                      & \multirow[c]{3}{*}{\begin{tabular}[c]{@{}c@{}}$30<\We_g<63$\\ (Multimode)\end{tabular}} & \multirow[c]{3}{*}{\begin{tabular}[c]{@{}c@{}}$40<\We_g<80$\\ (Plume-shear)\end{tabular}} & \multirow[c]{3}{*}{\begin{tabular}[c]{@{}c@{}}$41<\We_g<80$\\ (Multimode)\end{tabular}} & \multirow[c]{3}{*}{\begin{tabular}[c]{@{}c@{}}$\We_g>31$\\ (Sheet-thinning)\end{tabular}} & \begin{tabular}[c]{@{}c@{}}$45<\We_g<65$\\ (Bag-plume)\end{tabular}  \\ \cline{6-6}
                                      &                                                                                    &                                                                                      &                                                                                    &                                                                                      & \begin{tabular}[c]{@{}c@{}}$65<\We_g<85$\\ (Multi-bag)\end{tabular}  \\ \hline
Shear-stripping breakup               & $\We_g>63$                                                                           & $\We_g>80$                                                                             & $\We_g>80$                                                                           &                                                                                      & $\We_g>85$                                                           \\
\Xhline{4\arrayrulewidth}
\end{tabular}%
}
\end{table}

In this study, the breakup regimes of viscous droplets, especially the intermediate breakup regime, are classified in detail through the experimental morphology, and the transitions of different breakup regimes are obtained by theoretical analysis. The rest of the paper is organized as follows. The experimental setup is described in Section \ref{sec:sec2}. The results are presented and discussed in Section \ref{sec:sec3}, including the droplet morphology during the deformation and breakup, the theoretical analysis of the droplet deformation and the Rayleigh-Taylor instability with the droplet viscosity taken into account, and the regime map of droplet breakup in the $\We_g$--$\Oh$ space. Conclusions are finally drawn in Section \ref{sec:sec4}.

\section{Experimental setup}\label{sec:sec2}
\begin{figure}[t]
  \centering
  \includegraphics[scale=0.7]{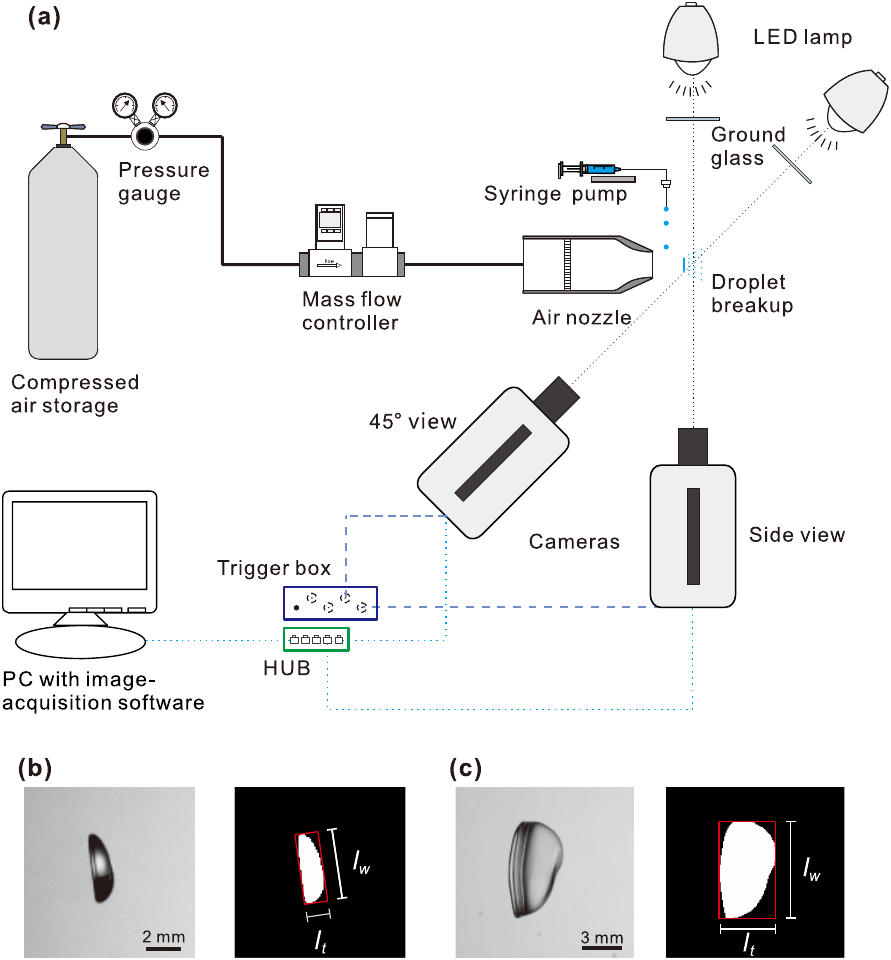}
  \caption{(a) Schematic diagrams of the experimental setup. (b-c) Illustration of the image post-processing for measuring the width ($l_w$) and thickness ($l_t$) of the droplet during the initial deformation stage and the bag development stage.}
\label{FIG:1}
\end{figure}

We used a continuous jet setup to study the droplet breakup, similar to previous studies  \citep{Jackiw2021InternalFlow, Xu2022ShearBreakup, Zhao2021ShearThickening, Xu2020ShearFlow}. The experimental setup is shown schematically in Fig.\ \ref{FIG:1}a. Droplets fell vertically into a horizontal jet and then broke up. The airflow of the jet was generated from a compressed-air cylinder, and the flow rate was controlled by a mass flow controller (Alicat MCRQ, maximum flow rate 3000 slpm, estimated uncertainty $\pm$ 0.8 \% of reading and $\pm$ 0.2 \% of full scale). Before ejected from a nozzle, the airflow was regulated by a honeycomb and two-layer mashes to reduce turbulence. The nozzle outlet is rectangular, 20 mm in width and 30 mm in height. To minimize the effect of the jet boundary layer, the falling velocity of the droplet was adjusted to ensure that the droplet deformed and broke up mainly in the core area of the jet at all jet velocities considered in this study. The experiments were performed at room temperature (25 $^{\circ}$C), and the density of the air is ${\rho _g} = 1.2$ kg/m$^3$. The range of the jet velocity ($u_g$) is 8.3–70 m/s. Test liquids are silicone oils of different viscosities, whose physical properties are shown in Table \ref{tab:tab2}. The droplet diameter used in this study is ${d_0} = 2.65 \pm 0.05$ mm. The ranges of the Weber number ($\We_g$) and the Ohnesorge number ($\Oh$) are also shown in Table \ref{tab:tab2}.

Images of the droplet breakup process were captured by two high-speed cameras. In addition to the side-view imaging, we also shot simultaneously from a 45$^{\circ}$ view to obtain more information about the morphology of the windward side of the droplet. The frame rate of the synchronized cameras was 10,000 frames per second (fps.), and the spatial resolution is 65–100 $\mu$m/pixel. Two macro lenses (Nikon AF 60 mm f/2.8D) with a small aperture (F16) were used to obtain an adequate field of view. Two high-power (280 W) light-emitting diode (LED) lights diffused by ground glasses were used as background light sources, which ensures sufficient brightness for the high-speed imaging.

After obtaining the raw images from the cameras, we post-processed the images using Matlab to quantify the shape evolution of the droplet, including the width ($\l_w$) and thickness ($l_t$) of the droplet. The shape of the droplet was obtained by converting the images into binary according to the brightness. Then the image processing is slightly different in the two stages of droplet deformation, i.e., the initial deformation stage and the piercing development stage of the droplet. In the initial deformation stage, considering that the droplet is flattened and may be slightly tilted, we determined the smallest bounding box covering the droplet. The long side of the bounding box is the droplet width, and the short side is the droplet thickness, as shown in Fig.\ \ref{FIG:1}b. After the droplet is flattened, the long and short sides of the smallest bounding box do not correspond to the droplet length and width due to the bag development or the periphery retraction. Therefore, during the piercing development stage, we fixed the bounding box vertically. Then, the horizontal length of the bounding box is regarded as the droplet thickness, and the vertical length is regarded as the droplet width, as shown in Fig.\ \ref{FIG:1}c. Finally, the image processing stops when the bag starts to break up.

\begin{table}
\centering
\caption{Physical properties of silicone oils and corresponding experiment conditions.}
\begin{tabular}{p{2.2cm}p{2.2cm}p{3cm}p{2.2cm}p{2.2cm}}
\toprule
Viscosity & Density & Surface tension & $\Oh$ & $\We_g$ \\
($\mu_d$, mPa$\cdot$s) &($\rho_d$, kg/m$^3$) &($\sigma$, mN/m) & &\\
\midrule
10   & 930		& 20.1		& 0.045   & 10--488 \\
20	& 950		& 20.6		& 0.088   & 12--476\\
50	& 960		& 20.8		& 0.22    & 13--472\\
100	& 966		& 20.9		& 0.43    & 16--734\\
200	& 968		& 21		& 0.86    & 22--730\\
500	& 971		& 21.1		& 2.15    & 36--727\\
\bottomrule
\end{tabular}
\label{tab:tab2}
\end{table}

\section{Results and discussion}\label{sec:sec3}
\subsection{Droplet morphology during the deformation and breakup}\label{sec:sec31}
\subsubsection{Transition from bag to bag-stamen breakup}\label{sec:sec311}
\begin{figure}
  \centering
  \includegraphics[scale=0.6]{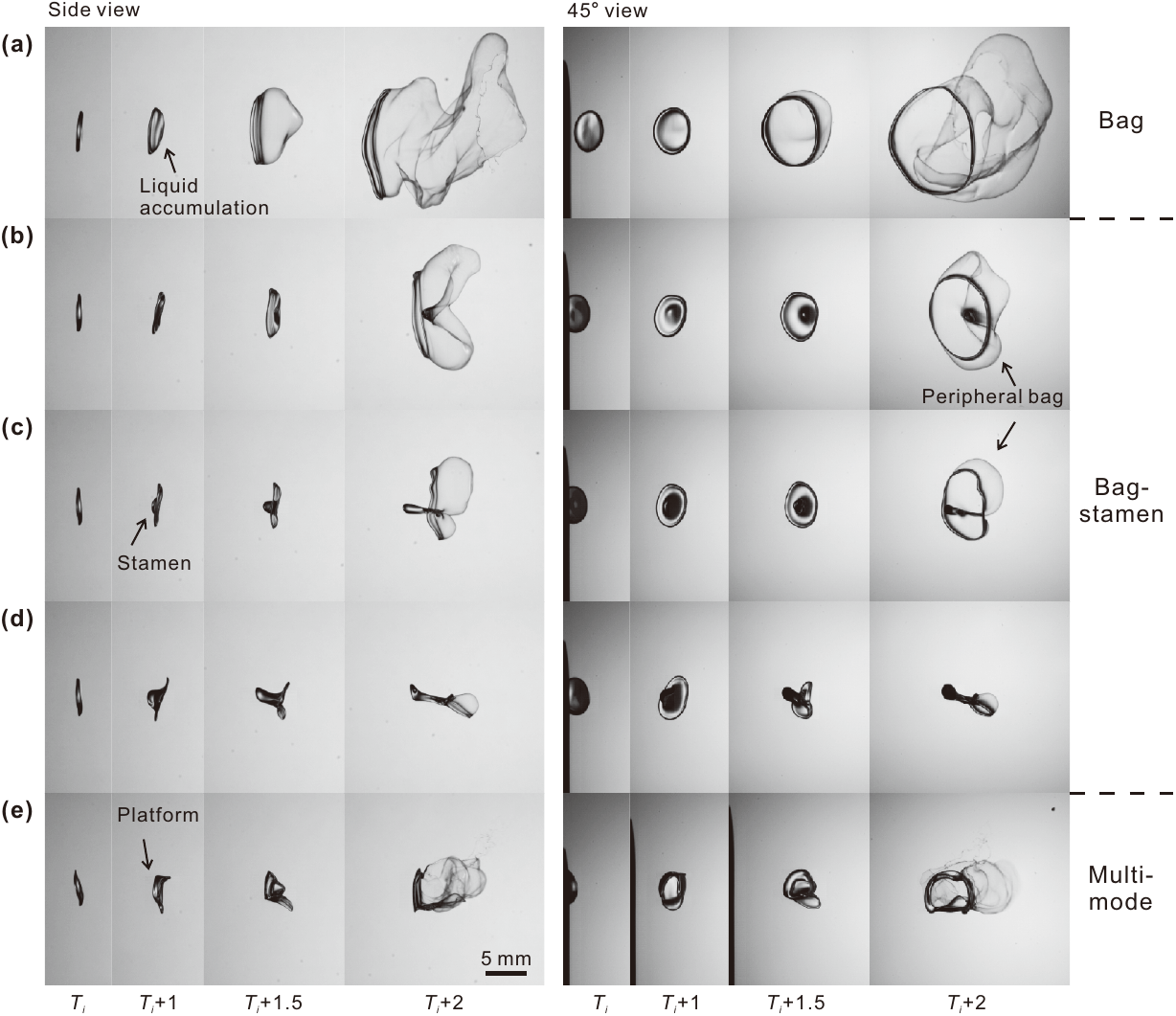}
  \caption{Synchronized image sequences from the bag regime to the multimode regime. The Weber numbers are $\We_g=31.7$ (a), 35.5 (b), 42.3 (c), 52.7 (d), and 75.1 (e), respectively. The Ohnesorge number is fixed at $\Oh=0.43$. $T_i$ is the dimensionless initial time for the droplet flattening. The corresponding movie can be found in the supplementary material (Movie 1).}
\label{FIG:2}
\end{figure}

The transition from the no-breakup to the bag breakup has been investigated by many previous studies \cite{Theofanous2012ViscousLiquids, Cohen1994ViscousBreakup, Gelfand1996DropletBreakup} and their results are similar to our experimental result. Hence, we start with the transition from the bag breakup to the bag-stamen breakup. Fig.\ \ref{FIG:2} shows the change of the droplet morphology after initial flattening under different $\We_g$ and the same $\Oh$. The time corresponding to the images is nondimensionalized by ${t^ * } = {d_0}\sqrt {{\rho _d}/{\rho _g}} /{u_g}$ for comparison.

The key feature for the transition from the bag regime to the bag-stamen regime is the liquid accumulation in the middle of the droplet as the bag develops. For the bag breakup at relatively higher $\We_g$ (close to the bag-stamen breakup regime), there is a slight accumulation of liquid in the middle of the droplet at the early stage of the bag development ($T_i+1$ in Fig.\ \ref{FIG:2}a, where $T_i$ is the dimensionless initial time for the droplet flattening). This induces a lower local stretching speed of the liquid film, then a slight depression on the bag ($T_i+1.5$ in Fig.\ \ref{FIG:2}a), which is similar to the dimple formation at the bag tip observed by Jackiw and Ashgriz \cite{Jackiw2021InternalFlow}. After that, the accumulated liquid may develop into a stamen via the drainage of liquid from the liquid film due to the surface tension effect. When the liquid viscosity is low, the accumulated liquid develops into a stamen quickly through rapid drainage. In contrast, when the viscous effect is not negligible, the drainage process is retarded, and the accumulated liquid is gradually consumed during the stretching of the liquid film ($T_i+2$ in Fig.\ \ref{FIG:2}a).

As $\We_g$ increases, the liquid accumulation in the middle of the droplet increases. The droplet breakup enters the bag-stamen regime when the accumulated liquid is enough to form a stamen. The formation of the stamen in the middle transforms the hollow bag in the bag breakup regime into a peripheral bag, as shown in Fig.\ \ref{FIG:2}b. As $\We_g$ further increases, the stamen gradually changes from a small liquid clump to a large liquid column. In contrast, the peripheral bag gradually becomes smaller until it becomes a sheet, as shown in Figs.\ \ref{FIG:2}b-d. Finally, the liquid accumulated in the middle of the droplet further increases to form a platform. When the platform is pierced, the droplet enters the multimode regime, as shown in Fig.\ \ref{FIG:2}e. Among these processes, as the bag grows in size, it can become asymmetric due to the asymmetry of the lateral development of the RT instability.

Liquid accumulation in the middle of the droplet is important for the droplet morphology. Jackiw and Ashgriz \cite{Jackiw2021InternalFlow} found that the breakup morphology is related to the division of the droplet into a disk that forms on the windward face and an undeformed droplet core. The amount of the droplet liquid contained in the undeformed core determines the breakup morphology. It should be mentioned that the mechanism of the liquid accumulation mentioned in this work is different from the theory by Jackiw and Ashgriz \cite{Jackiw2021InternalFlow}. In the work of Jackiw and Ashgriz \cite{Jackiw2021InternalFlow}, the `undeformed core' can be regarded as the accumulated liquid. In contrast, derived from our experimental images (shown in the first column of Fig.\ \ref{FIG:2}), we assume the droplet is almost completely flattened before the bag development. In addition, even with a similar degree of initial deformation, the liquid accumulation in the droplet middle is completely different (shown in the fourth column of Fig.\ \ref{FIG:2}). Therefore, we argue that the bag development after the initial deformation has an important effect on the liquid accumulation. After the initial flattening of the droplet, the liquid in the middle of the droplet accumulates as the bag develops, which is attributed to the squeezing effect caused by the lateral development of the bag and the drainage of liquid from the liquid film. The junction of two instability waves is prone to induce liquid accumulation.

\begin{figure}
  \centering
  \includegraphics[scale=0.5]{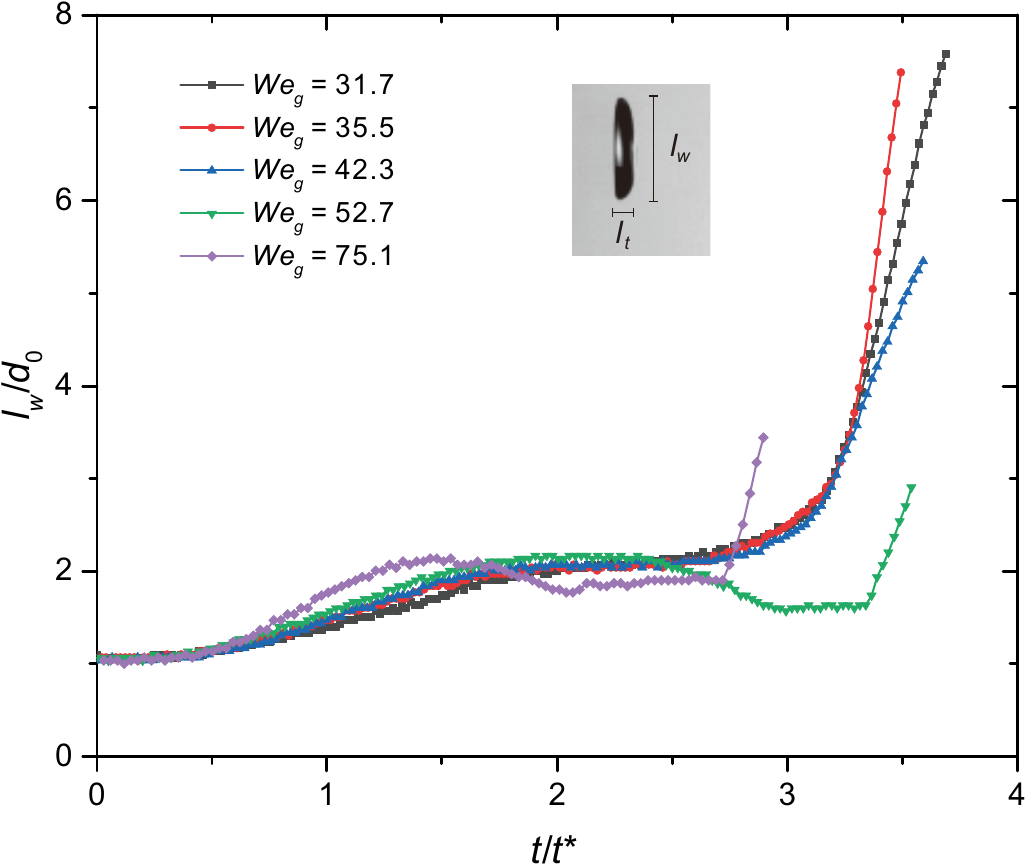}
  \caption{Evolution of the droplet width corresponding to Fig.\ \ref{FIG:2}. The slow increase in $l_w/d_0$ corresponds to the flattening of the droplet, and the rapid increase corresponds to the development of the bag.}
\label{FIG:3}
\end{figure}

To quantify the droplet shape, the evolution of the droplet width corresponding to Fig.\ \ref{FIG:2} is shown in Fig.\ \ref{FIG:3}. In the early stage ($t/{t^*}$ = 0--1.25), the droplet width increases as the droplet is flattened. For low $\We_g$ ($\We_g$ = 31.7, 35.5, 42.3), a bag develops after the droplet is flattened, so the droplet width continues to increase ($t/{t^*}$ = 1.25--3.5). But for high $\We_g$ ($\We_g$ = 52.7, 75.1), most of the liquid is accumulated in the middle of the droplet, while the periphery only contains a small amount of liquid. The mass differences cause the droplet to deflect after being flattened. Therefore, the droplet width first decreases to a certain extent after being flattened, and then increases as the bag develops. In particular, due to the droplet deflection before the bag develops, the rapid increase in $l_w/d_0$ at $\We_g$ = 52.7 is later than that at $\We_g$ at = 42.3.

\subsubsection{Transition from bag-stamen to multimode breakup}\label{sec:sec312}

As discussed in Section \ref{sec:sec311}, the amount of accumulated liquid in the middle of the droplet increases with increasing $\We_g$ until the piercing occurs. At the same time, the bag on the periphery weakens to become a sheet with increasing $\We_g$. During the transition from the bag-stamen to the multimode breakup regime, two scenarios will appear as $\We_g$ increases depending on the liquid viscosity.

\begin{figure}
  \centering
  \includegraphics[scale=0.75]{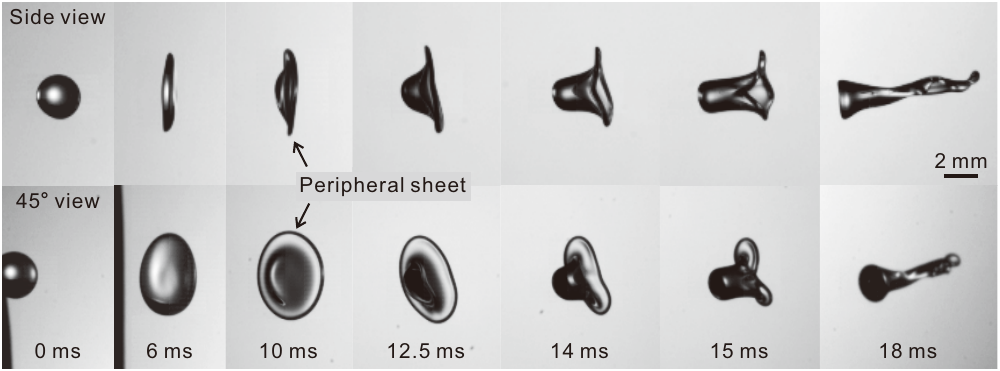}
  \caption{Synchronized images of the droplet retraction in the bag-stamen regime. Here, $\We_g$ = 62.3, $\Oh$ = 0.86. The corresponding movie can be found in the supplementary material (Movie 2).}
\label{FIG:4}
\end{figure}

\begin{figure}
  \centering
  \includegraphics[scale=0.75]{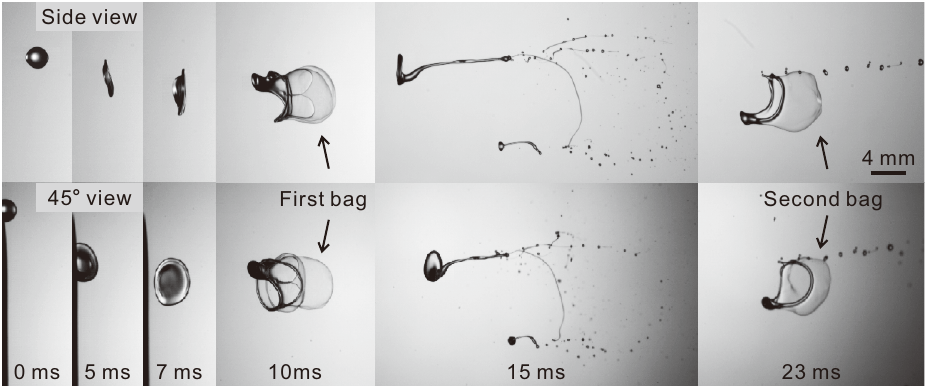}
  \caption{Synchronized images of the dual-bag morphology in the multimode regime. Here, $\We_g$ = 34.7, $\Oh$ = 0.088. The corresponding movie can be found in the supplementary material (Movie 3).}
\label{FIG:5}
\end{figure}

In the first scenario, for a high-viscosity droplet, the piercing in the middle of the droplet still does not occur after the periphery bag weakens to become a sheet. The sheet retracts toward the wake region behind the droplet, and no breakup is observed, as shown in Fig.\ \ref{FIG:4}. The morphology of the droplet retraction is similar to the wing formation observed by Radhakrishna \emph{et al.}\ \cite{Radhakrishna2021HighOh}. In the second scenario, for a low-viscosity droplet, the piercing in the middle of the droplet occurs before the periphery bag weakens to become a sheet. In this scenario, the dual-bag breakup can be observed, i.e., piercing occurs in both the middle and peripheral parts, as shown in Fig.\ \ref{FIG:5}. More details of the dual-bag breakup for low-viscosity droplets can be found in Ref.\ \cite{Cao2007DualbagBreakup}.

\begin{figure}
  \centering
  \includegraphics[scale=0.5]{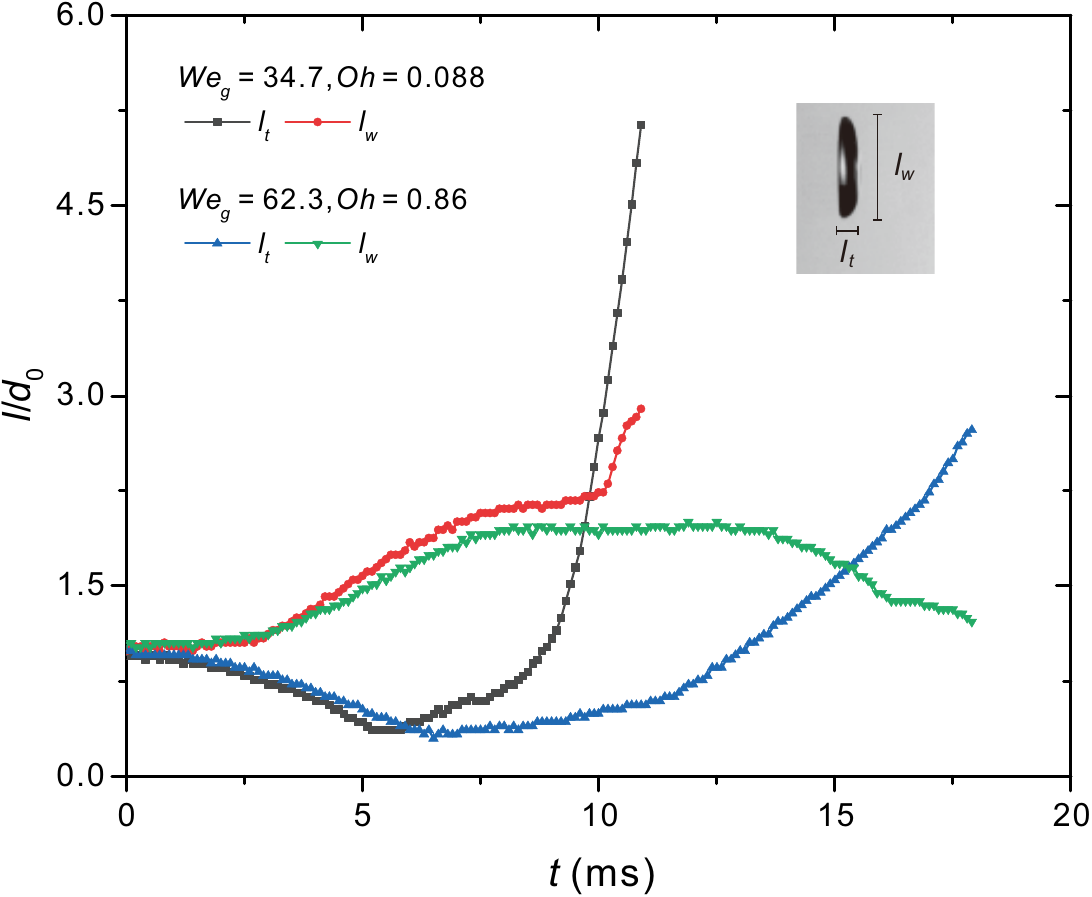}
  \caption{Evolution of the width and the thickness of the droplets corresponding to Figs.\ \ref{FIG:4} and \ref{FIG:5}.}
\label{FIG:6}
\end{figure}

To quantify the shape evolution of the droplet, the width and the thickness of the droplets corresponding to Figs.\ \ref{FIG:4} and \ref{FIG:5} are shown in Fig.\ \ref{FIG:6}. In the early stage (0--5 ms), the droplet width increases and the droplet thickness decreases as the droplet is flattened in both cases. For the low-viscosity liquid, the droplet is pierced rapidly after being flattened, so the width and the thickness of the droplet increase rapidly (5--12 ms). For the high-viscosity liquid, although the degree of flattening in the early stage is similar to that for the low-viscosity liquid, the development of the piercing is much slower (6--12 ms), thus allowing the droplet to retract completely before the piercing appears. As the droplet retracts, the droplet width decreases and the droplet thickness increases more slowly than in the low-viscosity case (12--18 ms).

Here, we clarify the dual-bag breakup and the droplet retraction in our classification of the breakup regimes. For the cases with different viscosities, the change of the morphology in the middle and the periphery of the droplet is different. Since the middle part of the droplet generally contains more liquid, whether the piercing occurs in the middle of the droplet has a greater impact on droplet atomization than the deformation at the periphery. Therefore, we take the change of the morphology in the middle of the droplet as the main criterion for the classification of different breakup regimes. In the case of the droplet retraction, no piercing occurs in the middle part of the droplet, and the resulted atomization effect is close to the bag-stamen breakup at high $\We_g$, so the droplet retraction is classified into the bag-stamen breakup regime. In contrast, in the case of the dual-bag breakup, the piercing in the middle of the droplet occurs, so the dual-bag breakup is classified into the multimode breakup regime. The stage at which the piercing occurs after the initial flattening is important for the accurate classification of breakup regimes. Moreover, considering that the droplet breakup is an integral process, in addition to using the middle deformation as the main criterion, we also examine the morphological changes in other stages, like the peripheral retraction. This criterion will be used throughout the paper.

\subsubsection{Multimode breakup}\label{sec:sec313}

The morphology of the multimode breakup is more complex than the bag-stamen breakup. If only using the side view like in previous experiments \cite{Dai2001MultimodeBreakup, Radhakrishna2021HighOh}, it is difficult to distinguish the complex morphology of the multimode breakup. Here, by combining the side view and the ${45^\circ}$ view, we have a clear understanding of the breakup morphology, which is important in determining the regime of breakup. According to the criterion of the morphology change in the middle of the droplet, we further classify the multimode breakup into the low-order multimode regime and the high-order multimode regime.

\begin{figure}[tbh]
  \centering
  \includegraphics[scale=0.75]{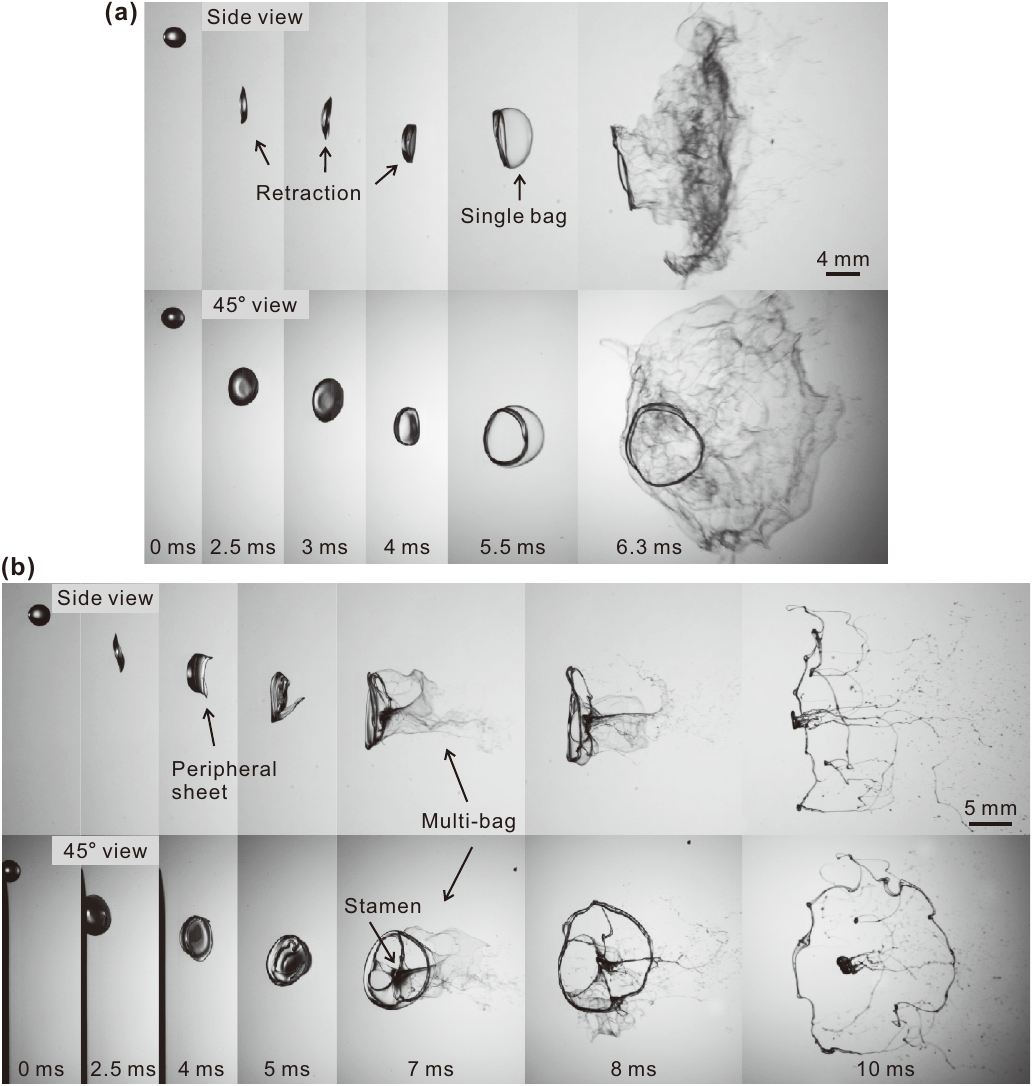}
  \caption{Synchronized images in the low-order multimode regime. (a) The single-bag structure at $\We_g$ = 356, $\Oh$ = 2.15, (b) the bag-stamen structure at $\We_g$ = 108, $\Oh$ = 0.43. The corresponding movie can be found in the supplementary material (Movie 4).}
\label{FIG:7}
\end{figure}

For the low-order multimode regime, the middle part of the droplet is pierced once after the sheet retraction or the bag breakup at the periphery. This single piercing can form multiple morphologies in the middle of the droplet. For the low-order multimode regime at low $\We_g$, the piercing causes the formation of a single-bag morphology in the middle of the droplet, as shown in Fig.\ \ref{FIG:7}a, which is similar to the rebound observed by Theofanous \emph{et al.}\ \cite{Theofanous2012ViscousLiquids}. For the low-order multimode regime at high $\We_g$, the piercing leads to the formation of multiple bags and a central stamen in the middle of the droplet, as shown in Fig.\ \ref{FIG:7}b. The transition of these morphologies in the middle of the droplet is similar to the transition from the bag regime to the bag-stamen regime discussed in Section \ref{sec:sec311}. The bag can be dramatically deformed in the late stage of bag development for high-viscosity droplets. This is because the high viscosity allows the liquid film to be stretched dramatically without breakup. In contrast, for low-viscosity liquids, the bag breaks up quickly into small droplets, as shown in Ref.\ \cite{Jackiw2022SizeDistribution}. The generation of film instability and the action of the airflow turbulence also causes the thin film to twist and entangle.

\begin{figure}
  \centering
  \includegraphics[scale=0.8]{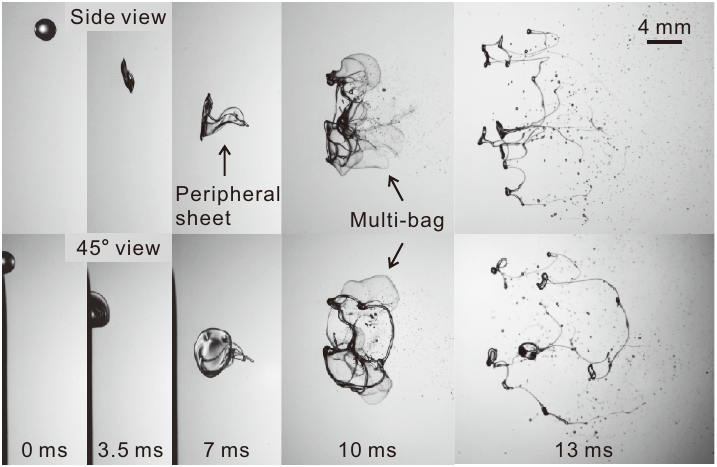}
  \caption{Synchronized images of the multi-bag structure in the low-order multimode regime. Here, $\We_g$ = 55.1, $\Oh$ = 0.088. The corresponding movie can be found in the supplementary material (Movie 5).}
\label{FIG:8}
\end{figure}

In the low-order multimode regime, $\Oh$ also affects the droplet morphologies. At the periphery of the droplet, at high $\Oh$, the droplet tends to form a complete peripheral sheet and then retract, while at low $\Oh$, the droplet tends to form a small bag or a single sheet. In the middle of the droplet, at high $\Oh$, the transition in the morphology with $\We_g$ is slow due to the viscous retardation, so the breakup can have the single-bag and bag-stamen structures, as shown in Fig.\ \ref{FIG:7}. In contrast, at low $\Oh$, the transition of different morphologies is very sensitive to $\We_g$, and the position of the bags formed by the piercing is random, as shown in Fig.\ \ref{FIG:8}. Previous studies tend to regard the low-order multimode breakup in low $\Oh$ cases as the random piercing of multiple bags, i.e., the multi-bag breakup \cite{Jain2015SecondaryBreakup, Jackiw2021InternalFlow}. However, no matter whether it is a single bag, bag-stamen, or multi-bag in the middle of the droplet, the piercing occurs altogether. Therefore, these morphologies are classified as the low-order multimode regime in this study.

\begin{figure}
  \centering
  \includegraphics[scale=0.75]{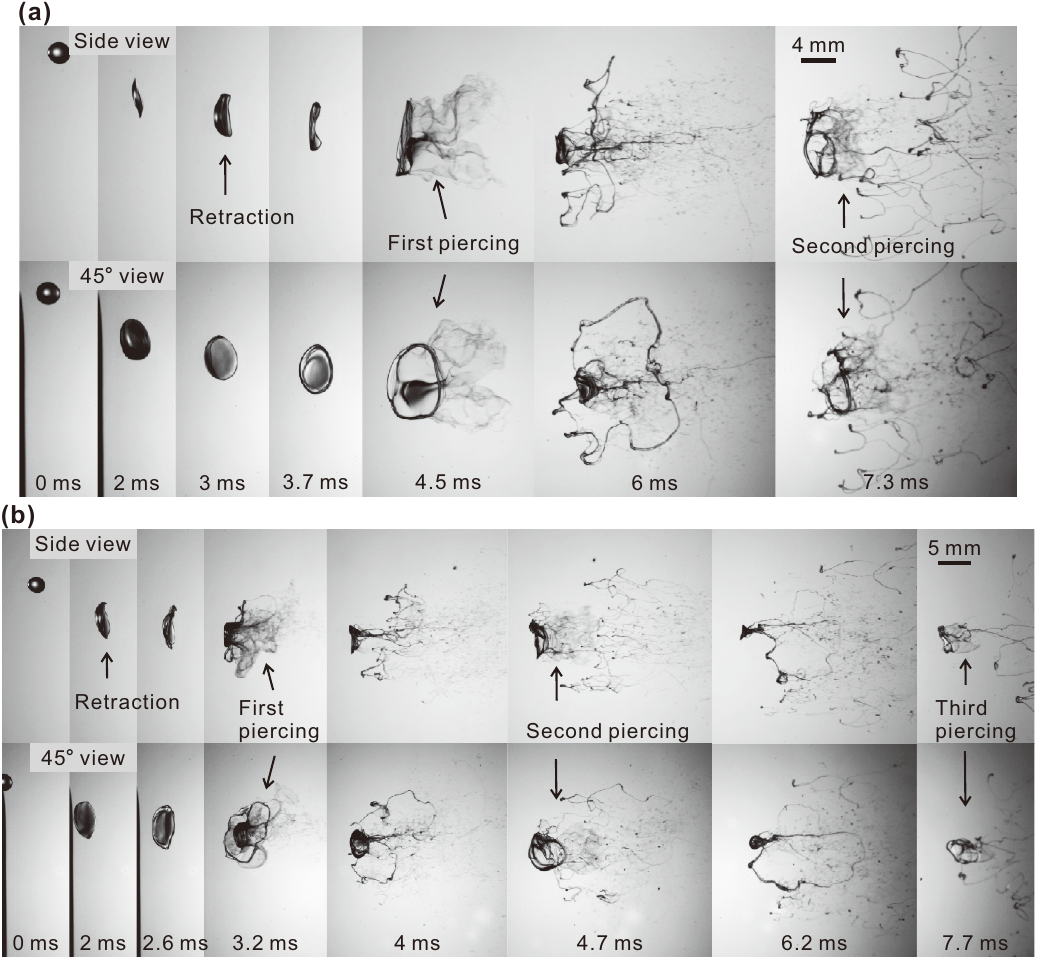}
  \caption{Synchronized images in the high-order multimode regime. (a) The double-piercing morphology at $\We_g$ =263, $\Oh$ = 0.86, (b) the triple-piercing morphology in the high-order multimode regime at $\We_g$ =358, $\Oh$ = 0.86. The corresponding movie can be found in the supplementary material (Movie 6).}
\label{FIG:9}
\end{figure}

The high-order multimode regime is characterized by multiple piercings in the middle of the droplet, and occurs at a higher $\We_g$ than the low-order multimode regime. The morphologies of double and triple piercing in the high-order multimode regime are shown in Fig.\ \ref{FIG:9}. The multiple piercings occur sequentially from the periphery to the middle of the droplet as time goes by, and each piercing contains multiple bags. In Fig.\ \ref{FIG:9}a, the middle of the droplet is first flattened and widened by the airflow (3--3.7 ms), and then slightly narrowed due to the squeezing effect caused by the lateral development of the bag during the first piercing (3.7--4.5 ms), and finally widened during the second piercing (4.5--7.3 ms). Similarly, the squeezing effect on the middle of the droplet caused by the lateral development of the bag also occurs in Fig.\ \ref{FIG:9}b. These indicate the importance of bag development in the liquid accumulation of the droplet middle.

The morphology of the multiple piercings in the middle of the droplet is similar to the repeated breakup described in the work of Jackiw and Ashgriz \cite{Jackiw2021InternalFlow} and the work of Dorschner \emph{et al.}\ \cite{Dorschner2020TransverseRTinstability}. But different from the theory of the undeformed core proposed by Jackiw and Ashgriz \cite{Jackiw2021InternalFlow}, we argue that the piercing controlled by the RT instability after droplet deformation determines the droplet morphology. The development of multiple RT instability waves across the width of a flattened droplet leads to the multiple breakup sequences, i.e., the multiple piercings. Therefore, based on this view, we further classify the multimode breakup into a low-order multimode regime and a high-order multimode. For the low-order multimode regime, the piercings in the middle of the droplet occur altogether, while for the high-order multimode regime, the piercings in the middle of the droplet occur sequentially.

\subsubsection{Shear-stripping breakup}\label{sec:sec314}

\begin{figure}
  \centering
  \includegraphics[scale=0.75]{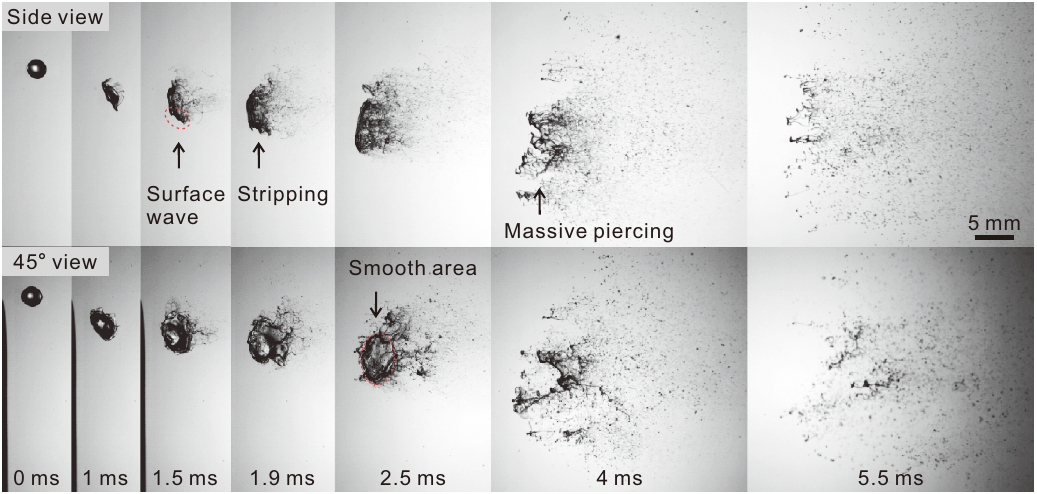}
  \caption{Synchronized images of the shear-stripping breakup. Here, $\We_g$ = 175.8, $\Oh$ = 0.045. The corresponding movie can be found in the supplementary material (Movie 7).}
\label{FIG:10}
\end{figure}

With increasing $\We_g$, the times of piercing increase and the liquid consumption of each piercing decreases. It can be inferred that these multiple piercings from the periphery to the middle of the droplet will gradually transform into stripping. The main feature of the transition to the shear-stripping regime is the appearance of Kelvin-Helmholtz (KH) instability waves at the periphery of the droplet (1.5 ms in Fig.\ \ref{FIG:10}). The KH instability waves originate from the shearing effect of the airflow at the periphery and can induce mass stripping (1.9 ms in Fig.\ \ref{FIG:10}). Compared with the piercing along the flow direction, the stripping has an outward trend (from the pole to the equator on the windward side of the droplet). This causes the windward side of the droplet to be tensioned. Therefore, even in a strong airflow, the windward side of the droplet still has a relatively smooth area (2.5 ms in Fig.\ \ref{FIG:10}). This phenomenon is consistent with the observation by Theofanous \cite{Theofanous2011DropBreakup}. However, in the later stage, the main body of the droplet is flattened and perpendicular to the direction of the airflow. Then, the droplet is pierced by the airflow in the later stage (4 ms in Fig.\ \ref{FIG:10}).

\subsection{Regime map}\label{sec:sec32}
The droplet has different morphologies in different breakup regimes, which directly affect the breakup time and the fragment size in practical applications \cite{Pilch1987BreakupTime}. In addition, in many numerical simulations of atomization processes, the determination of the breakup regime is the first step when choosing from different numerical models, while the misjudgment of the breakup regimes may lead to huge errors \cite{Michele2019SpraySystems}. Therefore, it is important to determine the transition conditions of different breakup regimes.

\begin{figure}
  \centering
  \includegraphics[scale=0.8]{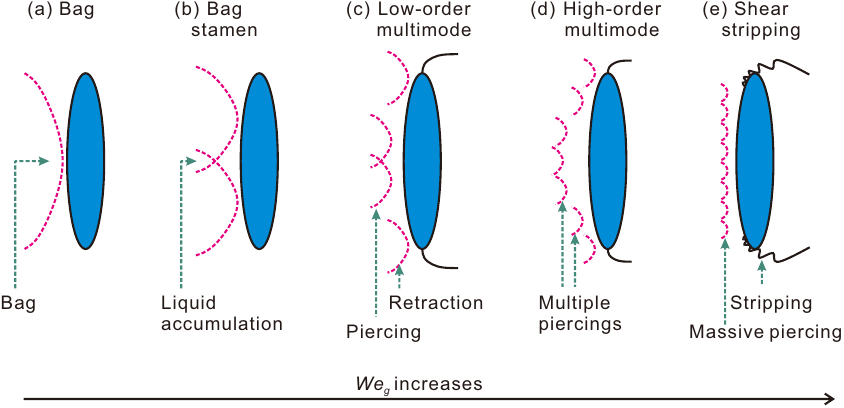}
  \caption{Illustration of the relationship between the RT instability waves and the droplet morphology in different regimes. The dashed lines indicate RT instability waves.}
\label{FIG:11}
\end{figure}

For different breakup regimes, the piercing in the middle of the droplet after the initial flattening plays a critical role, as suggested by the droplet morphologies discussed in Section \ref{sec:sec31}. Previous studies have shown that the physical mechanism for the piercing is the development of the Rayleigh-Taylor (RT) instability waves on the windward side of the droplet \cite{Zhao2011BagBreakup, Theofanous2011DropBreakup, Theofanous2012ViscousLiquids, Asahara2022ShearRT}. Here, we use the RT instability analysis and extend it to the transition of the breakup regimes of viscous droplets, especially for the multimode breakup regime. The relationship between the RT instability waves and the droplet morphology in different regimes is shown in Fig.\ \ref{FIG:11}. As the airflow velocity increases, the number of RT instability waves on the windward side of the droplet increases, and different regimes appear in sequence. In the bag breakup regime, a single RT instability wave develops to form the bag morphology, as shown in Fig.\ \ref{FIG:11}a. In the bag-stamen regime, the wavelength of the RT instability wave decreases, and two RT instability waves develop. The junction of two instability waves is prone to induce liquid accumulation and then form a stamen at the middle of the droplet, as shown in Fig.\ \ref{FIG:11}b. In the low-order multimode regime, more RT instability waves develop, and the middle of the droplet is pierced. However, the development of the RT instability wave at the droplet edge is suppressed due to the rapid loss of relative velocity, which tends to cause the droplet edge retraction rather than piercing, as shown in Fig.\ \ref{FIG:11}c. In the high-order multimode regime, the number of the RT instability waves further increases. The slight difference in the thickness of the middle part of the droplet causes the multiple piercings to occur sequentially from the edge to the middle of the droplet, as shown in Fig.\ \ref{FIG:11}d. Finally, in the shear-stripping regime, the KH instability waves appear at the edge of the droplet, and the middle part of the droplet is pierced by many small waves in the later stage, as shown in Fig.\ \ref{FIG:11}e.

From the above analysis, we can see that the transition of different regimes is determined by the comparison between the droplet width after the initial flattening and the wavelength of the RT instability. Therefore, the transitions of the different regimes correspond to
\begin{equation}\label{eq:eq1}
  \frac{{{d_w}}}{{{\lambda _{RT}}}} = {C_1}\
\end{equation}
where ${d_w}$ is the droplet width after initial flattening, ${\lambda _{RT}}$ is the most-amplified wavelength of the RT instability, and $C_1$ is a constant and different for different transitions. This view was proposed by Zhao \emph{et al.}\ for the transitions between the no-breakup and the bag breakup \cite{Zhao2011BagBreakup} and between the bag-breakup and the bag-stamen breakup \cite{ Zhao2013BagStamen}.

It should be noted that both the surface tension and the viscosity affect the development of the RT instability. Particularly for high-viscous fluids, the viscous effect should not be neglected as shown in Section \ref{sec:sec31}. In this condition, the exact solution of the RT instability wavelength is difficult to obtain \cite{Mikaelian1993ViscousRT}. By assuming that the viscous and surface tension effects are additive to the leading order, Aliseda \emph{et al.}\ \cite{Aliseda2008ViscousAtomization} obtained a simplified form of the most-amplified wavelength of the RT instability for the atomization of a coaxial jet as
\begin{equation}\label{eq:eq2}
  {\lambda _{RT}} = 2\pi \left[ {{{\left( {\frac{{3\sigma }}{{{\rho _d}a}}} \right)}^{1/2}} + {{\left( {\frac{{\mu _d^2}}{{\rho _d^2a}}} \right)}^{1/3}}} \right]\
\end{equation}
where $a$ is the acceleration of the liquid. Based on Eqs.\ (\ref{eq:eq1}) and (\ref{eq:eq2}), Zhao \emph{et al.}\ \cite{Zhao2011BagBreakup} obtain
\begin{equation}\label{eq:eq3}
  {\left( {\frac{1}{{{\We_c}}}} \right)^{1/2}} + {C_2}{\left( {\frac{{{d_w}}}{{{d_0}}}} \right)^{1/3}}{\left( {\frac{{{\Oh^2}}}{{{\We_c}}}} \right)^{1/3}} = {C_3}
\end{equation}
where $\We_c$ is the critical Weber number for the transition between different regimes, ${C_2} = {\left( {{C_D}/36} \right)^{1/6}}$ and ${C_3} = {\left( {1/{\We_{c,0}}} \right)^{1/2}}$. When $\Oh$ approaches zero, $\We_c$ is equal to the critical Weber number in the inviscid case ($\We_{c,0}$). ${C_D}$ is the drag coefficient for the droplet and can be considered as a constant for simplicity.

\begin{figure}
  \centering
  \includegraphics[scale=0.5]{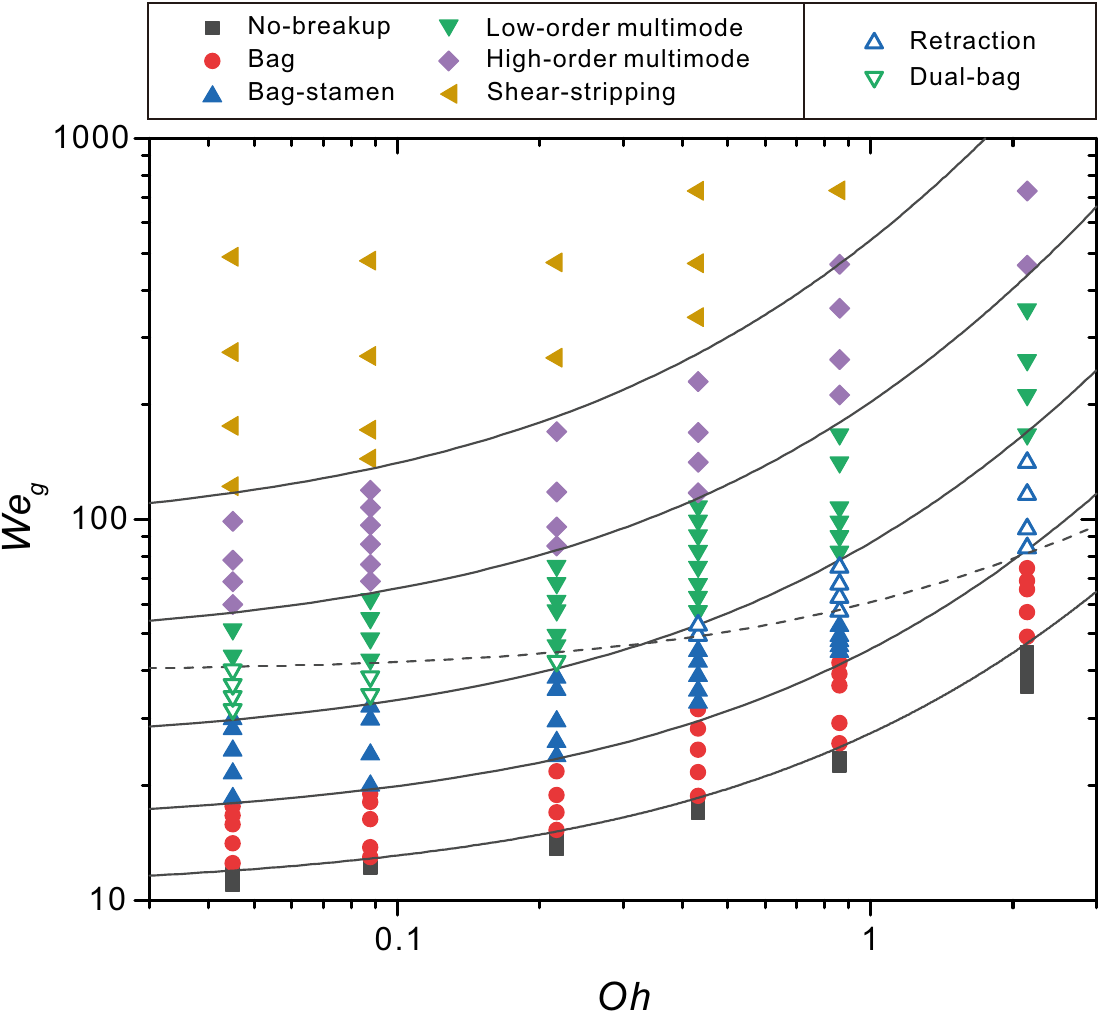}
  \caption{Regime map of the droplet breakup. The solid lines are based on the theoretical model [Eq.\ (\ref{eq:eq5})] for the transitions of different regimes. $C$ in Eq.\ (\ref{eq:eq5}) is 0.24 obtained by fitting. The critical Weber numbers in the inviscid case $\We_{c,0}$ are 10.5, 15.5, 25, 46, and 90, for the transitions from no-break to bag, bag-stamen, low-order multimode, high-order multimode, and finally shear stripping, respectively. The dashed line is based on Eq.\ (\ref{eq:eq6}) and corresponds to the boundary of the dual-bag breakup in the low-order multimode regime and the droplet retraction in the bag-stamen regime.}
\label{FIG:12}
\end{figure}

Here, to expand the applicability of the mode to the transitions of other breakup regimes, we further modify the model of Zhao \emph{et al.}\ \cite{Zhao2011BagBreakup}. In the inviscid case, the kinetic energy transferred by the airflow ($ \sim {\rho _g}d_0^3u_g^2$) is converted to the surface energy of the droplet ($ \sim \sigma d_w^2$). This yields
\begin{equation}\label{eq:eq4}
  \frac{{{d_w}}}{{{d_0}}} \sim \We_{c,0}^{1/2}
\end{equation}
At the same transition, the droplet has similar deformation morphology in the viscous and inviscid cases. Hence, the increase in the surface energy of the droplet in the viscous and inviscid cases should be the same. When the viscosity is present, the increase in $\We_c$ compared with the inviscid case is attributed to the extra viscous dissipation within the droplet \cite{Cohen1994ViscousBreakup}. Therefore, the deformation in the viscous case can also be described by Eq.\ (\ref{eq:eq4}). Through Eq.\ (\ref{eq:eq4}), we can simplify Eq.\ (\ref{eq:eq3}) to
\begin{equation}\label{eq:eq5}
  {\left( {\frac{{{\We_{c,0}}}}{{{\We_c}}}} \right)^{1/2}} + C\We_{c,0}^{2/3}{\left( {\frac{{{Oh^2}}}{{{\We_c}}}} \right)^{1/3}} = 1\
\end{equation}
where $C$ is a constant and can be considered as a weight coefficient of the viscous effect relative to the surface tension in the droplet breakup process. As shown in the regime map in Fig.\ \ref{FIG:12}, the transitions of different regimes can be well predicted by this model. To reduce the uncertainty of experimental results in Fig.\ \ref{FIG:12}, we obtain about seven breakup processes for each condition. The breakup processes are classified based on the criteria described in Section \ref{sec:sec31}, and the mode that has a higher occurrence rate is determined to be the breakup mode under the corresponding condition. Especially for conditions close to the transitions, the density of data points is increased to better identify the transitions of different regimes.

\begin{figure}[tb]
  \centering
  \includegraphics[width=0.98\columnwidth]{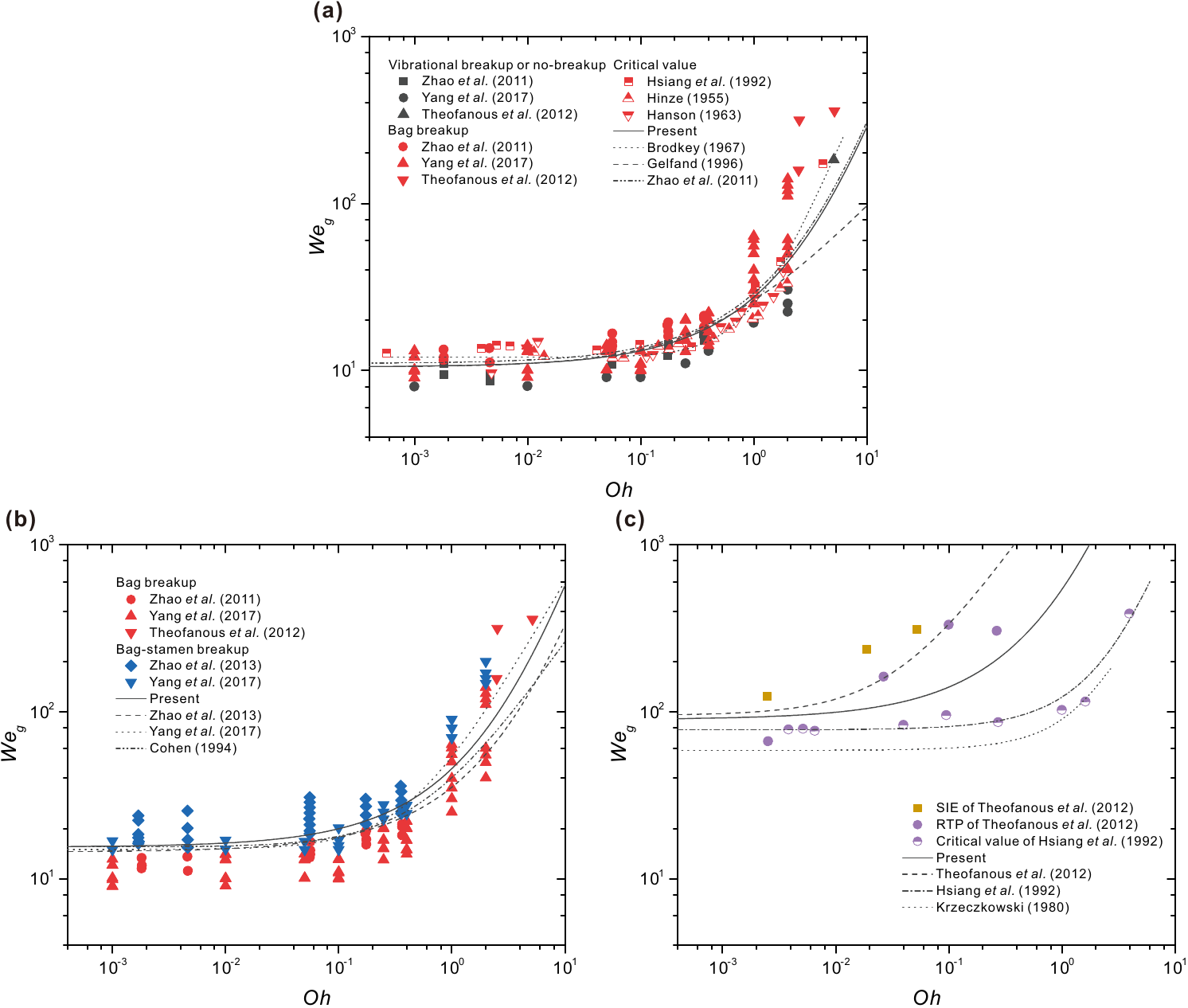}
  \caption{Comparison of our theoretical analysis [Eq.\ (\ref{eq:eq5})] with previous studies. (a) Transition from the no-breakup to the bag breakup. (b) Transition from the bag breakup to the bag-stamen breakup. (c) Transition from the multimode breakup to the shear-stripping breakup. The symbols are experimental results, and the lines are theoretical or empirical models.}
\label{FIG:13}
\end{figure}

In addition, this theoretical model is also compared with the results of previous studies. In previous studies, the theoretical or empirical models mainly focus on the transitions from the no-breakup to the bag breakup \cite{Zhao2011BagBreakup, Theofanous2012ViscousLiquids, Yang2017TransitionsWeberNumber, Hsiang1992SecondaryBreakup, Gelfand1996DropletBreakup, Hinze1955SheetThinningMechanism, Hanson1963VisousDroplet, Brodkey1967ViscousDroplet}, from the bag breakup to the bag-stamen breakup \cite{Zhao2011BagBreakup, Yang2017TransitionsWeberNumber, Zhao2013BagStamen, Cohen1994ViscousBreakup}, and from the multimode breakup to the shear-stripping breakup \cite{Theofanous2012ViscousLiquids, Krzeczkowski1980ViscosityBreakup, Hsiang1992SecondaryBreakup}. The comparison of the present model with those studies is shown in Fig.\ \ref{FIG:13}, and it shows good agreement with previous experimental data. In particular, comparing our model with the model of Zhao \emph{et al.}\ \cite{Zhao2011BagBreakup, Zhao2013BagStamen}, the difference between the two models for the transition from the no-breakup to the bag breakup is marginal (see Fig.\ \ref{FIG:13}a), while the difference in the transition from the bag breakup to the bag-stamen breakup is remarkable (see Fig.\ \ref{FIG:13}b). The difference between the two models originates mainly from the difference in the estimation of the degree of droplet deformation. The relationship for the extent of deformation used in the model of Zhao \emph{et al.}\ \cite{Zhao2011BagBreakup} is a semi-empirical equation for the maximum cross-stream diameter measured by Hsiang and Faeth \cite{Hsiang1992SecondaryBreakup}. It should be noted that the maximum cross-stream diameter of the droplet is measured at a certain moment. In the cases of bag breakup and bag-stamen breakup, the maximum cross-stream diameter generally corresponds to the moment of RT instability onset, so the semi-empirical equation is valid for the RT instability analysis. However, for the breakup regime at a high $\We_g$, the periphery of the droplet has retracted before piercing, and the piercings may occur at different moments, which makes the maximum cross-stream diameter based on the measurements at a certain moment invalid for the instability analysis. In our model, we estimate the cross-stream size of the droplet deformation based on the more general hypothesis of energy conservation. The estimated size can be considered as a characteristic cross-stream size of the liquid where the piercing occurs at different moments, which is valid for the instability analysis even at a high $\We_g$. Therefore, our model can be further extended to the classification for the multimode regime. In addition, due to the simplicity and effectiveness in estimating droplet deformation, our model is simple and effective to cover the transitions of all breakup regimes.

When the droplet breakup changes from the bag-stamen regime to the low-order multimode regime, the morphologies of the dual-bag and the retraction are observed as shown in Section \ref{sec:sec312}. These phenomena were also found in previous studies \cite{Radhakrishna2021HighOh, Cao2007DualbagBreakup}, but their ranges of occurrence have never been determined. For the dual-bag breakup, the piercing in the middle of the droplet occurs before the periphery retracts. For the droplet retraction, the piercing in the middle of the droplet still does not occur after the periphery retracts. This process is controlled by both the middle piercing and the peripheral retraction. The piercing can be reflected by Eq.\ (\ref{eq:eq5}). However, whether the periphery retracts or not depends on the loss of relative velocity between the droplet and the airflow. Hsiang and Faeth indicated that the droplet viscosity could lead to the loss of the relative velocity between the droplet and the airflow \cite{Hsiang1995BreakupCategory}. Using an average drag coefficient (${\bar C_D}$) and integrating the acceleration over the time interval from the initial moment to the breakup start moment, they obtained
\begin{equation}\label{eq:eq6}
  {\We_\text{local}} = \frac{{{\We_c}}}{{{{\left( {1 + {C_4}{{\Oh}}{{\We_c^{-1/2}}}} \right)}^2}}}\
\end{equation}
where ${\We_\text{local}}$ is the local Weber number of the droplet after the initial deformation, and ${C_4} \sim {\bar C_D}{({\rho _g}/{\rho _d})^{1/2}}$. Hsiang and Faeth viewed the droplet as a whole with a constant ${\bar C_D}$, and used Eq.\ (\ref{eq:eq6}) to predict the transitions for different deformation rates as well as the transition for the bag breakup.

For the regime transitions in our work, the middle and the periphery of the droplet need to be considered separately because of the piercing at the middle of the droplet and the retraction at the periphery. For the droplet periphery, if considering only the droplet periphery as a whole, the relative-velocity-loss model is still valid. The loss of relative velocity controls whether the droplet periphery retracts or not. Therefore, the model is appropriate for the transition of the dual-bag breakup or the droplet retraction. Here, we find this model can predict the boundary of the dual-bag breakup or the droplet retraction with fitting parameters $\We_\text{local} = 40$ and ${C_4} = 1.8$ based on our data, as shown by the dashed line in Fig.\ \ref{FIG:12}. Similarly, if considering only the droplet middle as a whole, the model by fitting the appropriate coefficient can reflect the relative loss in the droplet middle. However, due to the smaller drag coefficient and the larger mass at the droplet middle, the loss of the relative velocity at the middle of the droplet is much smaller than that at the periphery. In addition, the relative-velocity-loss model ignores the effect of the droplet viscosity on the piercing after the initial deformation and cannot explain the appearance of the multiple piercings. Therefore, for the piercing in the middle of the droplet, we still use the RT instability analysis.

Combining the boundary of the dual-bag breakup or the droplet retraction [Eq.\ (\ref{eq:eq6}), the dashed line in Fig.\ \ref{FIG:12}] and the transition between the bag-stamen regime and the low-order multimode regime [Eq.\ (\ref{eq:eq5}), the solid line in Fig.\ \ref{FIG:12}], we can obtain the ranges of the dual-bag breakup and the droplet retraction (see the hollow symbols in Fig.\ \ref{FIG:12}). For the low-viscosity case, below the dashed line means that the relative velocity loss at the periphery is relatively small and will not inhibit the development of RT instability waves at the periphery of the droplet. Above the solid line means that the number of RT instability waves is enough to cause piercing in both the middle and periphery of the droplet. Therefore, the dual-bag breakup, where piercing occurs in both the middle and peripheral parts, appears in the range below the dashed line and above the solid line for the low-viscosity case. For the high-viscosity case, above the dashed line means that the relative velocity loss at the periphery is large to inhibit the development of RT instability waves at the periphery. Below the solid line means that the RT instability waves can only cause piercing at the periphery, but the piercing is inhibited. Therefore, the droplet retraction, where no piercing occurs, appears in the range above the dashed line and below the solid line for the high-viscosity cases.

\section{Conclusions}\label{sec:sec4}

In this study, by combining high-speed images from the side view and the ${45^ \circ }$ view, we perform a detailed classification of the breakup regime of viscous droplets. Based on the morphology change in the middle of the droplet, the droplet breakup is classified into no-breakup, bag breakup, bag-stamen, low-order multimode, high-order multimode, and shear-stripping breakup. The transitions of different regimes are discussed. In the transition from the bag regime to the bag-stamen regime, the liquid accumulation in the middle of the droplet gradually increases, and the viscosity retards the development of the accumulated liquid into a stamen. In the transition from the bag-stamen regime to the low-order multimode, due to the different morphological changes in the middle and the periphery of the droplet, the dual-bag morphology appears in the low-viscosity case, while the droplet retraction appears in the high-viscosity case. Moreover, according to the times of piercings in the middle of the droplet, we classify the multimode breakup into a low-order multimode regime and a high-order multimode regime. In the low-order multimode regime, various morphologies appear in the middle of the droplet, including bag, bag-stamen, and multi-bag. In the high-order multimode regime, the morphologies of double-piercing and triple-piercing appear. Finally, as the times of piercings increase, the high-order multimode regime transits to the shear-stripping regime.

The interaction of the deformed droplet with the RT instability waves determines the breakup regime. By analyzing the viscosity effect on the droplet deformation and the RT instability, the transition conditions for each breakup regime of the viscous droplets are obtained. The present model can cover the transitions of all regimes. In addition, for the dual-bag morphology in the low-order multimode regime and the droplet retraction in the bag-stamen regime, the corresponding ranges are determined by combining the RT instability analysis with the loss of the relative velocity between the droplet with the airflow.

Droplet breakup in airflow is a complex process, especially for viscous droplets. The present study mainly focuses on the transitions of different breakup regimes. There are many open questions in this area yet to be answered. For example, for a specific breakup regime, the breakup time and the fragment size require further investigation in the future. The studies of the viscous droplet breakup can provide physical insight into liquid atomization, which is useful for the design and optimization of relevant applications.

\section*{Supplementary Material}
Supplemental Material include seven movies can be found with this article online.

\section*{CRediT authorship contribution statement}
\textbf{Zhikun Xu:} Conceptualization, Methodology, Validation, Investigation, Formal analysis, Writing-Original Draft, Visualization. \textbf{Tianyou Wang:} Resources, Supervision. \textbf{Zhizhao Che:} Software, Resources, Data Curation, Writing-Review ${\rm{\& }}$ Editing, Project administration, Supervision, Funding acquisition

\section*{Declaration of Competing Interest}
The authors declare that they have no known competing financial interests or personal relationships that could have appeared to influence the work reported in this paper.

\section*{Acknowledgements}
This work was supported by the National Natural Science Foundation of China (Grant Nos. 51676137 and 51921004).

\bibliographystyle{elsarticle-num}
\bibliography{dropletBreakup}
\end{document}